\documentclass[a4paper,11pt]{article}
\pdfoutput=1 

\usepackage{jinstpub} 
 \usepackage[table,xcdraw]{xcolor}

\title{\boldmath An automated system to measure the quantum efficiency of CCDs for astronomy}

\author[a,1]{R. Coles,\note{Corresponding author.}}
\author[c]{J. Chiang,}
\author[b]{D. Cinabro}
\author[a]{J. Haupt}
\author[c]{H. Neal}
\author[a]{A. Nomerotski}
\author[a]{P. Takacs}

\affiliation[a]{Brookhaven National Laboratory,\\Upton, NY 11973, U.S.A.}
\affiliation[b]{Physics, Wayne State University,\\Detroit, MI 48201, U.S.A.}
\affiliation[c]{SLAC National Accelerator Laboratory,\\ Menlo Park, CA 94025, U.S.A.}

\emailAdd{rcoles@bnl.gov}

\abstract{We describe a system to measure the Quantum Efficiency in the wavelength range of 300 nm to 1100 nm of 40x40 mm n-channel CCD sensors for the construction of the 3.2 gigapixel LSST focal plane. The technique uses a series of instrument to create a very uniform flux of photons of controllable intensity in the wavelength range of interest across the face the sensor.  This allows the absolute Quantum Efficiency to be measured with an accuracy in the 7\% range. This system will be part of a production facility at Brookhaven National Lab for the basic component of the LSST camera.}


\proceeding{PRECISION ASTRONOMY WITH FULLY DEPLETED CCDS\\
  1 - 2 DECEMBER 2016\\
  BROOKHAVEN NATIONAL LABORATORY, U.S.A.}

\begin{document}
\maketitle
\flushbottom

\section{Large Synoptic Survey Telescope (LSST)}
\label{sec:intro}
\subsection{What LSST is Trying to Achieve}
\par The goals of the Large Synoptic Survey Telescope can be divided into four themes: taking
an Inventory of the Solar System, mapping the Milky Way, exploring the Transient Optical
Sky, and probing dark energy and dark matter. These goals have directed the technical
development of the telescope and its science requirements. Each area of sky will be visited
1000 times, for two 15s exposures in a given filter, over a ten year period. This will generate
temporal astrometric and photometric data on over 20 billion objects. A major challenge
over the next decade will be to gain an understanding of dark energy and dark matter by
using the LSST to obtain wide-field surveys of gravitational lensing, large-scale distribution
of galaxies, and light curves of an unprecedented number of supernova ~\cite{b}.
\par The LSST will image the entire visible sky every few nights, capturing changes that, after
10 years of observation, can be stitched together to create a time-lapse movie of the universe.
As the LSST generates images, it will process and upload that information for applications
outside of pure research. Platforms similar to Google Earth will build 3D virtual maps of
the sky that will be fully available to the public.

\subsection{The Focal Plane and Science Rafts}
\par The LSST focal plane will be made up of 21 Science Rafts that will each contain a $3\times3$
mosaic of sensors, for a total of 189 CCDs. LSST is purchasing CCDs from two vendors:
ITL and e2v, with each Science Raft being vendor-homogeneous. The LSST camera will be the
largest digital camera ever constructed. 
\par Each Science Raft will be mounted on a tower that holds 
the front-end electronics for read out, with each Science Raft capable of acting as an autonomous 
camera individually controlled via the Observatory Control System. The Science Rafts will be 
grouped on the focal plane, where their timing and control modules will be synchronized. 
The LSST Camera will sit in a cryostat  that includes: the final focusing lens of the LSST's 
optical system, a filter wheel system with five large optical filters, a filter change
system which inserts any one of the filters into the optical path, a shutter, and part of a
cryogenic system that maintains the focal plane at $100K$. The whole assembly is known as
the LSST Cryostat, which is about two meters in diameter, three meters long, and weighs
$6200lbs$.

\section{CCD Quantum Efficiency}
\label{sec:QE}
\par
\begin{figure}[htbp]
\centering 
\includegraphics[width=1\textwidth,origin=c,angle=0]{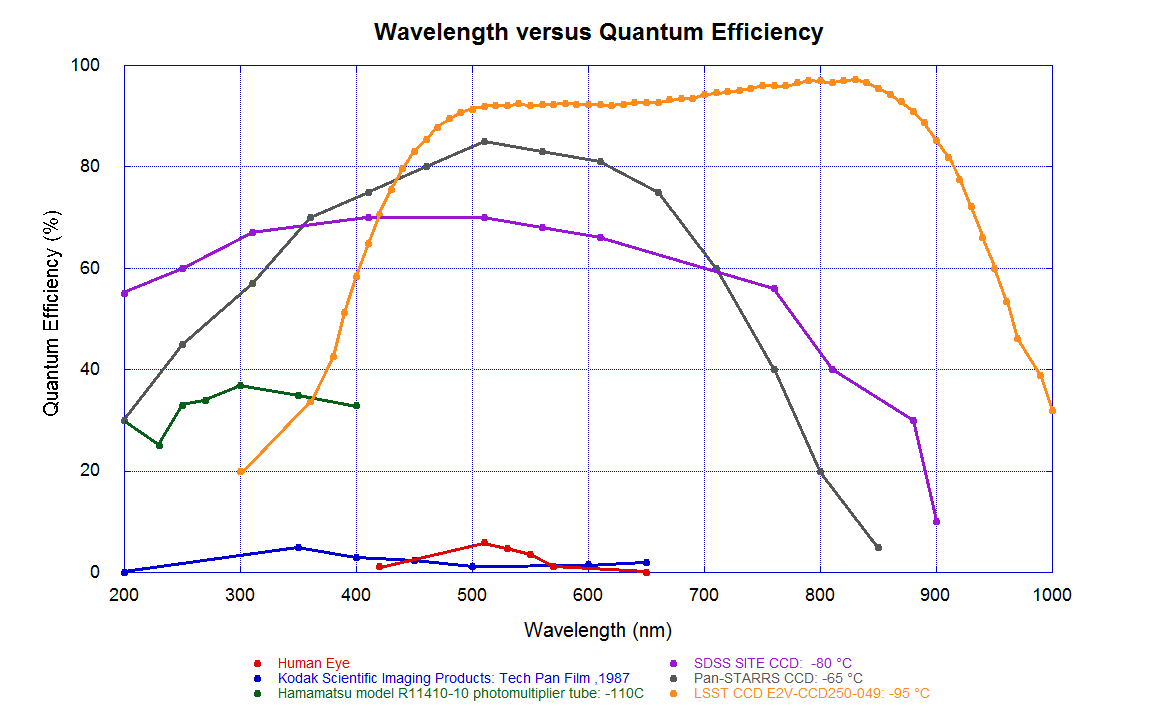}
\caption{\label{fig:i} QE of multiple devices: human eye ~\cite{c}, Kodak astronomical Tech Pan Film ~\cite{d}, Hamamatsu model R11410-10 photomultiplier tube ~\cite{e}, Sloan Digital Sky Survey (SDSS) CCD for photometric applications ~\cite{f}, Pan-STARRS CCD for photometric applications ~\cite{g}.}
\end{figure}

 A sensor's quantum efficiency (QE) is defined by its ability to convert photons into a
useful output. The human eye, for example, only has good QE between about $450 - 650nm$.
The QE of a CCD is defined as the fraction of photons incident on the CCD's surface that
are successful in creating electron-hole pairs, where the electrons are captured and measured
by the CCD's read out electronics, as shown in Equation \eqref{eq:x}.
\begin{equation}
\label{eq:x}
\begin{split}
QE &= \frac{N_{e^{-}}}{N_\gamma}
\end{split}
\end{equation}
where $N_{e^{-}}$ is the number of electrons generated and read out by the CCD, and $N_\gamma$ is the
number of photons incident on the CCD's surface. Figure~\ref{fig:i} shows the QE of multiple
devices ranging from the human eye to LSST CCDs.

\subsection{How Quantum Efficiency is Measured and Calculated}
\par Each pixel in a serial register is shifted out one at a time to the output electronics, where
the the total charge is converted to a voltage and then converted to a digital number, known
as an analog-to-digital-unit ($ADU$). In Equation \eqref{eq:x} we show the general form of the QE formula, but the form that is
used to analyze the electron-to-photon ratio in terms of the current output of a photodiode
in the light beam (photon measurement), and the current output of the CCD (electron
measurement), involves calculating the $ADU$ for each pixel:
\begin{equation}
\label{eq:y}
ADU_{pixel} = ADU_{exposed}  -  ADU_{dark}
\end{equation}
where $ADU_{exposed}$ is the analog-to-digital-unit per pixel measured by the CCD and its output
electronics during an exposure. Similarly, $ADU_{dark}$ is the ADU
during a dark exposure, and $ADU_{pixel}$ is the actual ADU per pixel when corrected for dark
current. These measurements are recorded in a Flexible Image Transport System file (FITS)
file that contains the ADU per pixel for every exposed pixel in the CCD ~\cite{h}. Thus,
Equation \eqref{eq:y} represents matrix subtraction. To find the number of collected and measured
electrons per pixel:
\begin{equation}
\label{eq:e}
e^{-}_{pixel}  = \frac{ADU_{pixel}}{g}
\end{equation}
where $g$ is the gain assigned to the A/D converter. To calculate the amount of photons
incident on the surface of the CCD, we create an electro-optical testing station that incidents
a uniform beam of light onto the CCD. To monitor the intensity of the incident light, a NIST
calibrated photodiode is put in the position where the CCD will eventually sit, and is exposed
to the beam. The ratio of the current from the photodiode, and a second diode at a different
location in the system, is used to calibrate the QE measurements to accurately represent the
number of photons incident on the CCD. The dark subtracted measurement of the number 
of photons incident on the surface of the NIST photodiode in the CCD position is:
\begin{equation}
\gamma_{pd} = \frac{I_{pd, exposed} - I_{pd, dark}}{E_{\gamma}A_{pd}R_{pd}(\gamma)}
\end{equation}
where $\gamma_{pd}$ is the number of photons incident on the NIST photodiode in the CCD position per square meter per second,
$I_{pd, exposed}$ is the current of the photodiode when exposed, $I_{pd, dark}$ is the photodiode current
during a dark exposure, $A_{pd}$ is the active area of the photodiode, $E_{\gamma}$ is the energy of the
incident photons, and $R_{pd}(\gamma)$ is the photodiode responsivity. The responsivity of the photodiode at a given wavelength is provided by the manufacturer and are in units of amp per Watt. To find the number of photons per pixel:
\begin{equation}
\label{eq:gamma}
\gamma_{pixel} = \gamma_{pd} A_{pixel}
\end{equation}
where $\gamma_{pixel}$ is the number of photons per pixel on the CCD per second, and $A_{pixel}$ is the area
of the CCD pixels. By combining Equation \eqref{eq:e} and Equation \eqref{eq:gamma} we calculate the QE for
each pixel in the CCD:
\begin{equation}
QE_{CCD} =  \frac{e^{-}_{pixel}}{\gamma_{pixel}t}
\end{equation}
where $t$ is the exposure time, and $e^{-}_{pixel}$ is the number of collected and measured electrons
per pixel as shown in Equation \eqref{eq:e}. Getting an accurate QE measurement is heavily dependent on
the uniformity of the flux incident on the measuring photodiode. To achieve such accuracy,
we construct a electro-optical system to deliver uniform light to the surface of a CCD that
is in vacuum and at its operating temperature, as shown in Figure~\ref{fig:j}.

\section{LSST Quantum Efficiency Test Stations}
\par
\begin{figure}[htbp]
\centering 
\includegraphics[width=1\textwidth,origin=c,angle=0]{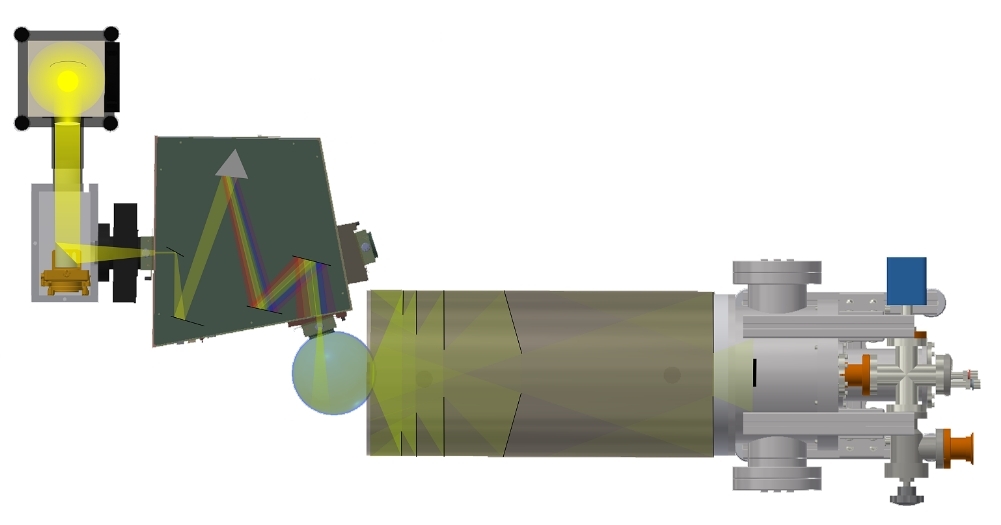}
\caption{\label{fig:j} The path of light through the LSST QE test station. Our light source is a  $300W$ xenon arc lamp. The light from the lamp is reflected off of an off-axis parabolic mirror, through a shutter and filter wheel. The filter wheel holds $305nm$ and $590nm$ cut-on filters. The filters help reduce stray light and avoid second-order effects from the monochromator. The Cornerstone 260 monochromator, with $1200lines/mm$ $200-1400nm$ range grating, uses the wavelength dispersion of the diffraction grating to filter light. The light exits the monochromator into a $6in$ diameter Labsphere integrating sphere. Since the sphere's surface illuminates isotropically, and the light is reflected multiple times, the light loses its spatial information and emerges as a uniform source. The light then enters a $590mm$ drift space. Since the distance that the light travels after emerging from the integrating sphere is proportional to its uniformity, and inversely proportional to its flux. The drift space distance was chosen to optimize uniformity versus sacrificed flux. The light leaving the drift space enters a Brookhaven National Laboratory custom designed cryostat, which holds the CCD is in vacuum and at its operating temperature of $-100ºC$.~\cite{a}}
\end{figure}

 The twin QE measurement stations will measure the QE of every sensor that we are considering for inclusion on the LSST focal plane. Figure~\ref{fig:j} is a model of the QE stations in the LSST clean room at Brookhaven National Laboratory.
\par To create uniform light, the system would has an enclosed $300W$ xenon arc lamp with an output flux that remains constant and has low drift. The lens inside of the lamp housing has been adjusted to focus the light onto the following off-axis parabolic mirror. The light from the lamp then reflects off of the mirror, which is aligned to focus the light through an open iris shutter, glass filter, and onto the motorized slits of a Cornerstone 260 monochromator. The iris shutter is used to regulate exposure times, and opens quickly so as not to create any non-uniformities that would occur from the shutter monetarily being in a partially open or closed state. The glass $305nm$ and $590nm$ bandpass filters block stray light and second-order effects from the monochromator. The monochromator uses the wavelength dispersion of a $1200lines/mm$ $200-1400nm$ range diffraction grating to filter light, allowing only the exact wavelength desired to enter the attached integrating sphere. The light in the $6in$ diameter Labsphere integrating sphere reflects enough times to ensure that the exiting beam has lost all spatial information and emerges as a plane wave. 
\par Since the uniformity of the light is partially dependent on the distance from the output port of
the sphere to the CCD, the light emerges from the integrating sphere into a drift space. 
The black flocking and baffles inside of the dark space would remove reflection, so the reflected light does not get more than
once chance to be absorbed by the CCD. The dark space is long enough for the light to be
become uniform and cover the entire area of the CCD with enough intensity to be
distinguishable from dark current. A cryostat, with a glass window, sits at the end of the dark space and keeps the CCD at optimal
operating temperature of $-100ºC$ and in vacuum at $10^{-6}-10^{-7}Torr$. 
\par The Camera Control System (CCS) software automates data acquisition and analysis for the test stations. Once the sensor is installed into the cryostat, the test station operator is free to leave the system while the CCS software handles the: pressure changes, cooling, measurements, and analysis. The average measurement run, from the installation of the sensor to post analysis removal, takes approximately 20 hours, with the measurements and analysis taking about 12 hours and the rest of the time dedicated to pressure and temperature changes.

\label{sec:Stations}
\subsection{Uncertainty Budget}
\par The LSST electro-optical (EO) stations are the main measurement systems for QE testing of the LSST sensors, and therefore our ability to accurately measure
the QE is dependent on the ability of the EO stations to have sufficient discrimination to
detect variation in the sensors. With this in mind, we put great effort into identifying the sources of uncertainty in our measurement process, quantify the uncertainties, and codifying their effects as reported values in an uncertainty budget. This involves studying
both calibration and measurement system capability. To create our uncertainty budget, we use the methods described in ~\cite{i}, as adapted to our needs.
\par We
combine the fractional uncertainties by root-sum-squares (quadrature) to obtain the standard
uncertainty, $u$, which is the standard deviation taking into account all the sources of
random and systematic uncertainty that affect the measurement result. The uncertainty for
each of the elements that we were unable to correct for using calibration or data analysis are
shown individually in Table~\ref{tab:j}. Here, the total expanded uncertainty is:
\begin{equation}
U = ku
\end{equation}
where $k$ is the critical value from the cumulative distribution function that is well known and
calculated by various groups such as NIST. For large degrees of freedom, $k = 1$ approximates
68\% confidence.

\section{Results}
\label{sec:Results}
\par
\begin{figure}[htbp]
\centering 
\includegraphics[width=1\textwidth,origin=c,angle=0]{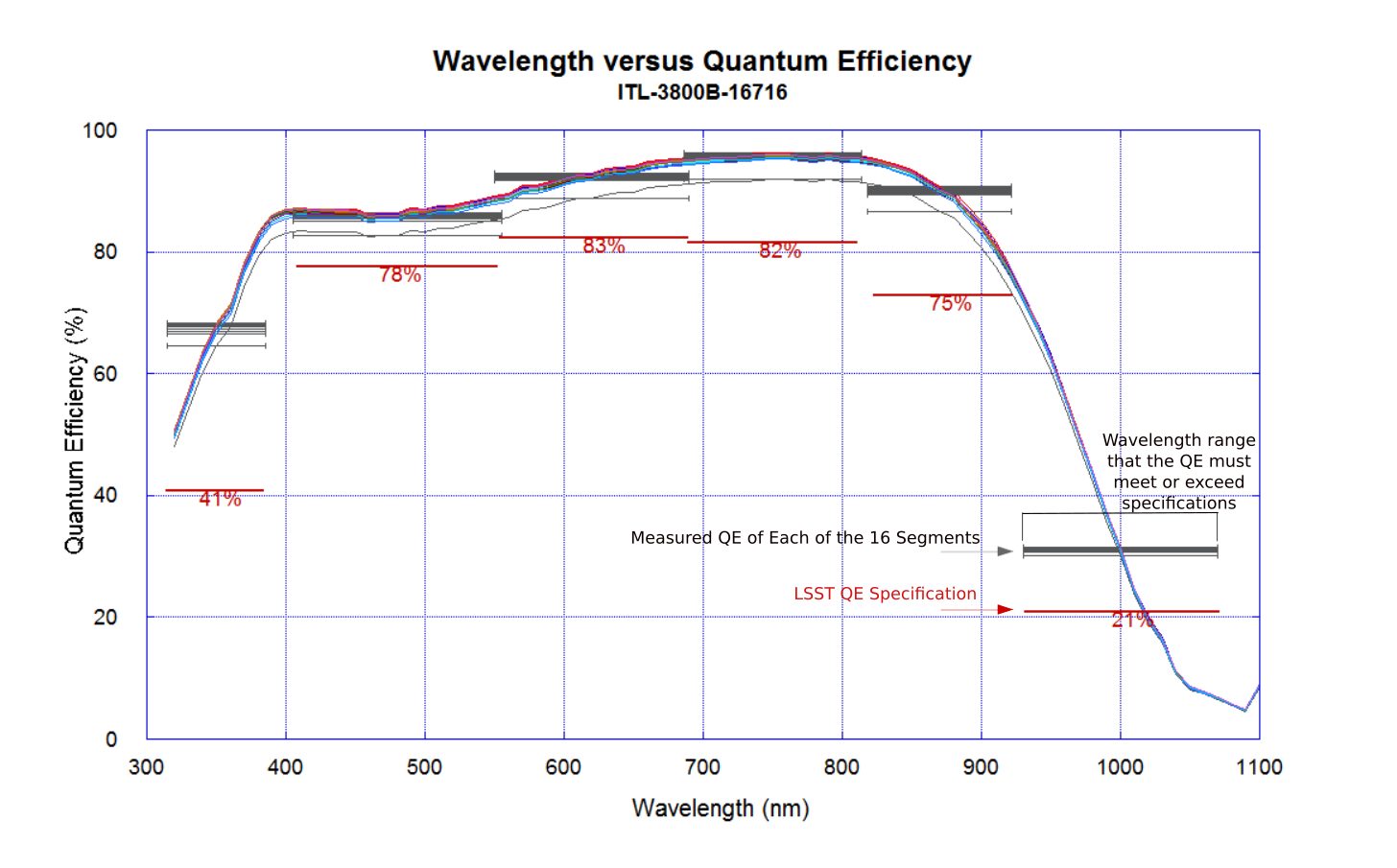}
\caption{\label{fig:m} Quantum efficiency curve for a LSST ITL CCD: this is an enhanced version
of the type of QE plots that we generate for our test reports. Each of the colored lines
represent the QE measured for each of the amplifiers of the CCD. The gray horizontal bars
represent the wavelength ranges at which the QE must exceed the LSST QE specifications.
The specifications for the QE at the six required wavelengths is shown here by the red
lines.}
\end{figure}

 The QE plots that we generate using our QE measurements stations typically look like
Figure~\ref{fig:m}. Though our specifications only require that we measure the QE at six wavelengths, we typically measure the CCDs every $10nm$ from $300 - 1100nm$, to check for any
abnormalities that may occur outside of the required wavelengths. Below $300nm$ we see very
low QE, as photons incident on the sensor at those wavelengths usually reflect off of the CCDs
surface or otherwise get absorbed in its surface layers. Light at wavelengths above $1100nm$
typically have an absorption length that is longer than the depth of silicon in the CCD,
making the sensor transparent to it. At wavelengths between approximately $500 - 800nm$
we see our best performance, and although no device is perfect, the QE in this range can
exceed 90\% for our sensors.
Figure~\ref{fig:m} is an enhanced version of the QE plots that we generate for our CCD test
reports. Here we show the QE for each of the 16 amplifiers individually, as opposed to the
average for the entire device. At this stage of development, we are very interested in studying
the uniformity of the QE across the sensor, and identifying segments that are preforming
poorly. The horizontal bars mark the QE at each wavelength that has an LSST specification
for performance. We ensure that each segment of the sensor meets the minimum specifications, as shown in Figure~\ref{fig:m} by the red lines. This sensor is currently being considered to
be used on our Engineering Test Unit (ETU) which will be our first engineering model of
our Science Raft.
\begin{table}[!h]
\centering
\begin{tabular}{|c|c|c|c|c|c|}
\hline
\rowcolor[HTML]{C0C0C0} 
\textbf{Measurement} & \textbf{Component} & \textbf{\begin{tabular}[c]{@{}c@{}}Sensitivity\\ Coefficient\end{tabular}} & \textbf{\begin{tabular}[c]{@{}c@{}}Mean \\ Measured\\ Value\end{tabular}} & \textbf{\begin{tabular}[c]{@{}c@{}}Standard \\ Deviation\end{tabular}} & \textbf{\begin{tabular}[c]{@{}c@{}}Fractional\\ Uncertainty\end{tabular}} \\ \hline
Reproducibility & \begin{tabular}[c]{@{}c@{}}LSST and \\ Vendor QE \\ Measurement\\ Stations\end{tabular} & $\sqrt{\frac{1}{3}}$ & \begin{tabular}[c]{@{}c@{}}47.3\%\\ (at 400nm)\end{tabular} & \begin{tabular}[c]{@{}c@{}}2.3\%\\ (maximum)\end{tabular} & 0.05 \\ \hline
Instrument Bias & Lamp Drift & 1 & 1.06 $\times10^{-10}$A & 2.28$\times10^{-12}$A & 0.02 \\ \hline
Instrument Bias & \begin{tabular}[c]{@{}c@{}}Glass \\ Cryostat\\ Window \\ Reflection\end{tabular} & 1 & 3.23\% & 0.13\% & 0.04 \\ \hline
Instrument Bias & Gain & 1 & 4.45e/ADU & 0.16e/ADU & 0.04 \\ \hline
Instrument Bias & \begin{tabular}[c]{@{}c@{}}NIST \\ Photodiode\\ Absolute \\ Calibration\end{tabular} & 1 & \begin{tabular}[c]{@{}c@{}}1.21A/W\\ (Relative \\ Expanded \\ Uncertainty \\ at 955nm)\end{tabular} & \begin{tabular}[c]{@{}c@{}}0.006A/W\\ (maximum)\end{tabular} & 0.005 \\ \hline
\rowcolor[HTML]{C0C0C0} 
Total &  &  &  &  & 0.067 \\ \hline
\end{tabular}
\caption{\label{tab:j} Uncertainty budget for LSST QE measurement system. Since our QE measurements scale linearly with gain, lamp intensity, etcetera, then the systematic fractional uncertainty in QE can be estimated as being the quadrature sum of the fractional uncertainties in this table.~\cite{a}. At the current time, the maximum uncertainty for the system is 6.7\%, though at most wavelengths the uncertainty is approximately 1-2\%. To create our uncertainty budget, we use the methods described in the \textit{NIST/SEMATECH e-Handbook of Statistical Methods} The sensitivity coefficient, $a$, shows the relationship of an individual component of an uncertainty to the standard deviation of the reported value. ~\cite{g}}
\end{table}

\newpage

\end{document}